\documentclass[aps,pre,preprint,eqsecnum,showpacs,draft,psfig]{revtex4} 
\usepackage{amsmath,bm}

\usepackage{psfig}

\begin{document}

\title{Casimir forces in modulated systems.}
 
\author{I.E.Dzyaloshinskii} \affiliation {Phys. Dept., University of
California, Irvine, UCI, CA 92697, USA}
 
\author{E.I.Kats}
\affiliation {Laue-Langevin Institute, F-38042, Grenoble, France } \affiliation {L. D. Landau Institute
for Theoretical Physics, RAS, 117940 GSP-1, Moscow, Russia.}

\date{\today}
 
\begin{abstract} 
For the first time we present analytical
results for the contribution of electromagnetic fluctuations
into thermodynamic properties of modulated
systems, like cholesteric or smectic
liquid crystalline
films. 
In the case of small dielectric anisotropy 
we have derived explicit analytical expressions for
the chemical potential of such systems.
Two limiting cases were specifically considered:
(i) the Van der Waals (VdW) limit, i.e., in the case when the retardation of
the electromagnetic interactions 
can be neglected; and (ii) the Casimir limit,
i.e. when the effects of retardation
becomes considerable.
It was shown that in the Casimir
limit, 
the film chemical potential oscillates with
the thickness of the film. 
This
non-monotonic dependence of the chemical potential on the film thickness
can lead to step-wise wetting phenomena, surface anchoring reorientation
and other important effects. 
Applications of the results may
concern the various systems in soft matter or condensed matter physics
with multilayer or modulated structures.                                                                                        
 
\end{abstract}
 
 
\maketitle
 
\section{Introduction} 
\label{int} 

Stratified systems are omnipresent in soft matter and solid state
physics.
Understanding of molecular interactions in these systems
is an important step for controlling their properties and
aggregation processes. Though these interactions may have
many different specific sources, the electromagnetic fluctuations
(VdW and Casimir forces) are their common cause.
These interactions become appreciable in the submicron range and rapidly
increase in the nanometer scale. Although, evidently,
the interactions would not scale nor ad infinitum, neither at shorter scales,
where steric short range interactions, i.e. interaction forces whose origin is in the Pauli principle,
become dominating ones. Still, submicron - nanometer range, where long range forces
are dominant ones, is the most important range for applications in nanotechnology processing.
Several approaches have been employed to calculate
these fluctuational forces. The simplest one is Hamaker
approach \cite{HA37}, summing all pairwise interactions,
which 
is evidently not applicable to condensed media, where
all molecular units are strongly correlated.
The rigorous continuum method was developed in \cite{DL61} 
(see also \cite{LL86}), by Lifshitz,
Pitaevskii, and one of the authors of the present paper (I.D.)
who derived the general
expression
for the electromagnetic fluctuational interaction between two macroscopic
isotropic bodies separated by an isotropic film.
Although the generalization of the analysis \cite{DL61} for
the anisotropic and multilayer systems is conceptually
straightforward, it deserves some attention, as it implies
prohibitively tedious and bulky calculations,
which could be done analytically only under certain rather
restrictive approximations
(see e.g., \cite{SZ01}, \cite{PP03}). Luckily, however, there are some cases
when these approximative calculations can be performed in a well
controlled way, and besides they may be useful for systems 
interesting for applications.
For example when dielectric anisotropy is small in
the whole region of frequencies relevant for the interactions,
one can use the regular perturbation theory \cite{kats}, which is a 
typical case for liquid crystalline films \cite{GP95}, \cite{CH92}.
Moreover, cholesteric liquid crystals having a long wavelength helical orientational modulation
can be treated
as a continuous model of multilayer systems.

In what follows we calculate the electromagnetic fluctuational contributions
to the chemical potential of a cholesteric or smectic liquid
crystal film (thickness $l$) confined
between two semi infinite isotropic media (having in mind
experimentally interesting cases, the latter can be, e.g., solid,
glass-like substrate and air). 
Our approach is a macroscopical one, i.e. we consider
quantities averaged over physically infinitesimal volumes.
Thus we restrict ourselves to only the long wavelength
part of the electromagnetic field fluctuations. In this range
everything can be expressed in terms of macroscopic
characteristics of the system (dielectric and magnetic
permeabilities).

Since for cholesteric liquid crystals typical scales of orientational
modulations are very large (of the order of $500 nm$ and even larger for nematic - cholesteric
mixtures \cite{GP95}, \cite{CH92}), they may be the perfect candidates for observing of
predicted (by our macroscopic
continuum approach) behavior. 
With a certain modification in order to take into account
magnetic properties, one can also perform analogous calculations for
large pitch magnetic spiral structures. 
The same is true for so-called lyotropic smectic liquid crystals where density
modulations occur at scales larger than characteristic molecular lengths $(2 - 5) nm$), however,
for thermotropic smectic liquid crystals this window of validity of our approach is much more narrow
(if it exists at all). In this case the continuum theory loses all pretense of quantitative
predictability and a complete description should incorporate microscopic
structure and molecular short range forces what can be done only numerically
(see e.g., \cite{BN02}, \cite{BN03}). 

Our motivation for presenting this discussion is two new predictions
which have emanated from our investigation. 
Namely, in the Casimir limit we have found spatially oscillating terms in the
chemical potential of the film. In the VdW
limit (small film thickness) we have found a monotonic in $l$ finite anisotropic correction to the Hamaker
constant, making possible a reorientation anchoring phase
transition with variation
of the film thickness.

The remainder of our paper has the following structure. 
Section \ref{fluc} contains basic methodical details
and equations necessary for our investigation in the frame work
of a regular method for calculating higher order
perturbative corrections. 
New results are discussed
in section \ref{chem} in subsections \ref{VdW} and  \ref{cas}, where we
present the electromagnetic fluctuational contributions into the
cholesteric film chemical potential.
Finally, Section \ref{con} deals with miscellaneous
subjects related
to the physical consequences of our results for wetting,
stability and anchoring phenomena in cholesteric films.
Some technical material is collected in
Appendix to the paper, and those readers who are not very interested in 
derivations can
skip this Appendix finding
all essential physical results in the main text of the paper.                                                   

\section{Electromagnetic fluctuations in cholesteric films}
\label{fluc}
We limit our analysis to the long range forces induced
by electromagnetic fluctuations, thus neglecting
all kinds of structural thermal fluctuations.
We consider a leftmost half space, $-\infty < x < 0$, as an isotropic body
(denoted in what follows by the indices 1 (solid, e.g., glass substrate)
and a rightmost half space 2, $l < x < + \infty $, (vacuum or air), 
separated by non-homogeneous and
anisotropic
film, $0 \leq x \leq l$,  
(denoted by the index 3) with thickness $l$.

First closely following the general method
\cite{DL61} we study the case when the substance 3 is a cholesteric
liquid crystal film. 
We chose the system for our benchmark to determine all specific features
of electromagnetic fluctuational forces in anisotropic modulated systems.
We take the $x$-axis perpendicular to the separating surfaces (planes
$x=0$ and $x=l$).
This is not the place to explain the general approach to fluctuational
electromagnetic forces in detailed, however, in a stripped down
version the theory \cite{DL61} is reduced to the calculation of Green
functions $D_{ik}$ to the Maxwell equations,
i.e.,
\begin{eqnarray}
\label{d1}
[\epsilon _{i l}({\bf {r}}, i|\omega _n|)\omega _n^2 - curl _{im} curl _{ml}]
D_{lk}({\bf {r}}, {\bf {r}}^\prime ; \omega _n) = - 4 \pi \delta ({\bf {r}} - {\bf {r}}^\prime )
\delta _{ik}
\, ,
\end{eqnarray}
where 
$\epsilon _{i l}({\bf {r}}, i|\omega _n|)$ is dielectric permeability 
at Matsubara imaginary frequencies $i \omega _n$. It is worth
noting that the dielectric function is always real on the imaginary axis
irrespective of whether the system is absorbing or not.
In what follows we use atomic units, i.e. put light velocity $c$ and Planck constant
$\hbar $ as 1 
(except where explicitely stated to the contrary
and the occurrences of $\hbar $ or $c$ are necessary for understanding).                                               

Calculating somehow $D_{ik}$ one can find the stress tensor
$\sigma _{ik}$ in the film 
\begin{eqnarray}
\label{d2}
\sigma _{ik}^\prime = -\frac{T}{2\pi } \sum ^{\infty }_{n=0} \left \{\epsilon ^{(3)}_{im}
\left [D_{mk}^E({\bf {r}},
{\bf {r}} ; i \omega _n ) - \frac{1}{2}\delta _{mk}D_{jj}^E({\bf {r}},
{\bf {r}} ; i \omega _n )\right ] - \frac{1}{2}\delta _{ik} D_{jj}^H({\bf {r}},
{\bf {r}} ; i \omega _n )\right \}
\, .
\end{eqnarray}
Here, $\sigma _{ik}^\prime $ is electromagnetic
fluctuation part of $\sigma _{ik}$, and functions $D_{ik}^E$ and $D_{ik}^H$ are corresponding
averages
of fluctuating electric and magnetic field components
\begin{eqnarray}
\label{d3}
D_{ik}^E({\bf {r}},
{\bf {r}^\prime } ; i \omega _n ) = - \omega _n^2 D_{il}({\bf {r}},
{\bf {r}^\prime } ; i \omega _n )
\, ; \,
D_{il}^H({\bf {r}},
{\bf {r}^\prime } ; i \omega _n ) = curl _{il} curl ^\prime _{km}
D_{l m}({\bf {r}} , {\bf {r}^\prime } ; i \omega _n )
\, .
\end{eqnarray}
Because the problem is homogeneous in $ z - y $ plane
we make a Fourier transform with respect to $z - z^\prime $ and
$y - y^\prime $ coordinates, thus getting $D_{ik}(x , x^\prime ,
{\bf {q}} ; i\omega _n)$. 
Taking into account that 
$\epsilon ^{(1)}_{ik} = \epsilon _1\delta _{ik}$,
$\epsilon ^{(2)}_{ik} = \delta _{ik}$, and
\begin{eqnarray} && 
\label{d4} 
\epsilon ^{(3)}_{ik} =
\left [ 
\begin{array}{ccc} 
\epsilon _3 & 0 & 0 \\  
0 & \epsilon _3 +\epsilon _a (1 + \cos(2 q_0 x))  
& \epsilon _a \sin (2 q_0 x)  \\ 
0 & \epsilon _a \sin (2 q_0 x)  & \epsilon _3 + \epsilon _a (1 - \cos(2 q_0 x))  
\\ 
\end{array} 
\right ] \, ,
\end{eqnarray}                                                                                                  
where we define the cholesteric dielectric
permeability as $\epsilon _{ik}^{(3)} = \epsilon _3 \delta _{ik} + 2 \epsilon _a
n_i n_k$ (${\bf {n}} = (0 \, , \, \cos (q_0 x) \, , \,  \sin (q_0 x))$ is cholesteric equilibrium
director, 
$q_0$ is cholesteric spiral modulation wave vector, related to cholesteric
pitch $p$ as $q_0 = 2\pi /p$,  and in what follows we assume a
small dielectric anisotropy $\epsilon _a <
\epsilon _3$, what is the case for all known cholesteric liquid crystals
\cite{GP95}, \cite{CH92}.

In the natural coordinate frame $x , y , z$ (\ref{d2}) is a set of nine
coupled equations (presented in the appendix to the paper).
This set should be supplemented by the usual boundary conditions (corresponding to the continuity of
the tangential components of the electric and magnetic fields) which are the
continuity conditions for 
\begin{eqnarray}
\label{d5}
D_{yk} \, , \, D_{zk} \, , \,
curl _{yl} D_{lk} \, , \, curl _{zl} D_{lk}
\, .
\end{eqnarray}
In principle the full information concerning electromagnetic
fluctuational forces is contained in solution of (\ref{a1}) - (\ref{a9}) 
supplemented by the boundary conditions
(\ref{d5}). In practice, however, for a given cholesteric
dielectric permeability (\ref{d4}) the analytic solution
is infeasible, thus some approximations are definitely
needed not only to make calculations possible, but also to understand
correctly underlying physical phenomena.
Luckily since $\epsilon _a <
\epsilon _3$, for all known cholesteric liquid crystals, one can solve
the equations using a regular perturbation theory with respect
to the small parameter $\epsilon _a/\epsilon _3$.

Omitting a large amount of tedious (although straitforward) algebra,
we get in the first order approximation the set of four equations to find
$D_{zz} \, , \, D_{yy} \, , \, D_{xy} $, and $D_{xx}$. Two latter equations
are simple relations
\begin{eqnarray}
\label{d6}
D_{xy} = - \frac{i q}{w^2}\frac{\partial }{\partial x}D_{yy}
\, , \,  D_{xx} = - \frac{ i q}{w^2} \frac{\partial }{\partial x}D_{xy} - \frac{4 \pi }{w^2}\delta (x -
x^\prime )
\, ,
\end{eqnarray}
where we designate $w^2 \equiv \epsilon \omega _n ^2 + q^2$, and
as $\epsilon $ one should take $1 \, , \, \epsilon _2 $ and $\epsilon _3 + \epsilon _a$,
for air, substrate, and cholesteric film respectively. 
Two other equations (for $D_{zz}$ and $D_{yy}$) deserve some precaution
only in the cholesteric film region ($0 < x < l$) where they are equivalent
to one particle Schr\"odinger equations in a periodic potential (see appendix,  (\ref{a10}),
(\ref{a11})).
However, since
to derive these equations
we already used the perturbation theory, we have to consider only so-called
almost free particle approximation (weak coupling limit), which gives for our case
(\ref{a12}), (\ref{a13}). 
Determining arbitrary constants $A_i\, , \, C_i$ entering these expressions
from the boundary conditions (\ref{d5}) and subtracting homogeneous system contribution,
we find finally
almost conventional \cite{DL61} expressions for the Green functions
\begin{eqnarray}
\label{d7}
D_{zz} = \frac{4\pi }{w_3 \Delta }\cosh [w_3(x - x^\prime )]
\, ,
\end{eqnarray}
where, however, $\Delta $ contains oscillating terms
\begin{eqnarray}
\label{d8}
\Delta  = 1 - \exp (2w_3 l)\frac{(w_3 + w_1)(w_3 + w_2)}{(w_3 - w_1)(w_3 - w_2)}[1 +
\alpha _1 \sin (2 q_0 l) + \alpha _2\cos (2 q_0 l)]
\, ,
\end{eqnarray}
and we designate
\begin{eqnarray}
\label{d9}
\alpha _1 = \frac{\epsilon _a \omega _n^2}{2(q_0^2 + w_3^2)}\left [
- \frac{w_3}{2 q_0} + \frac{q_0}{w_3 + w_2}\right ]
\, ,
\end{eqnarray}
and
\begin{eqnarray}
\label{d10}
\alpha _2 = \frac{\epsilon _a \omega _n^2}{4(q_0^2 + w_3^2)}
\, .
\end{eqnarray}
Analogously
\begin{eqnarray}
\label{d11}
D_{yy} = \frac{4\pi w_3}{\omega _n^2 \epsilon _3 \Delta _1}\cosh [w_3(x - x^\prime )]
\, ,
\end{eqnarray}
where
\begin{eqnarray}
\label{d12}
\Delta _1  = 1 - \exp (2w_3 l)\frac{(w_3 + w_2)(\epsilon _1 w_3 + w_1)}{(w_3 - w_2)(\epsilon _1w_3 -
w_1)}[1 +
\beta _1 \sin (2 q_0 l) + \beta _2\cos (2 q_0 l)]
\, ,
\end{eqnarray}
and as above
\begin{eqnarray}
\label{d13}
\beta _1 = \frac{\epsilon _a w_3^3}{4(q_0^2 + w_3^2)\epsilon _3 (w_3 + w_2)} +
\frac{\epsilon _a w_2 w_3 q_0}{4\epsilon _3 (q_0^2 + w_3^2)}
\, ,
\end{eqnarray}
and
\begin{eqnarray}
\label{d14}
\beta _2 = - \frac{\epsilon _a w_3^3}{4\epsilon _3(q_0^2 + w_3^2)(w_3 + w_2)}
- \frac{\epsilon _a q_0^2}{2\epsilon _3(q_0^2 + w_3^2)}
\, .
\end{eqnarray}
These expressions (\ref{d7}) - (\ref{d14}) are our main results 
and they are ready for further inspection.

\section{Fluctuational contribution to the chemical potential}
\label{chem}

Our aim is to calculate fluctuational contributions into thermodynamic properties
of a thin cholesteric film on an isotropic solid substrate. Since the film should be
in equilibrium
\begin{eqnarray}
\label{d15}
P = P_{ch} - \sigma _{xx}^\prime
\, ,
\end{eqnarray}
where $\sigma _{xx}^\prime $ is given in (\ref{d2}), $P$ is the air or vapor pressure,
and $P_{ch}$ is the pressure of the bulk cholesteric.
Note that using a standard definition for the pressure
\begin{eqnarray}
\label{d16}
P_{ch} = \rho \frac{\partial E}{\partial \rho } - E
\, ,
\end{eqnarray}
were $\rho $ is cholesteric mass density,
and into the energy density $E$, orientational deformation energy
(Frank energy \cite{GP95}, \cite{CH92}) has to be added to isotropic liquid
pressure $P_0$. Therefore,
\begin{eqnarray}
\label{d17}
P_{ch} = P_0 + K_{22} q_0 \gamma (\epsilon _{ikn} n_i \nabla _k n_n - q_0)
\, ,
\end{eqnarray}
where $\varepsilon _{ikn}$ is antisymmetric tensor, $K_{22}$ - twist orientational
elastic modulus, and $\gamma = - (\partial \ln q_0/\partial \rho )$ is
cholesteric pitch ''compressibility''.
Now we are in the position to reap fruits of our calculations.
From (\ref{d15}) and given above expressions (\ref{d2}), (\ref{d7}) and (\ref{d11})
the electromagnetic fluctuational part of the cholesteric film chemical potential $\mu (l)$
reads
\begin{eqnarray}
\label{d18}
\mu (l)  = -\frac{T}{2\pi }\sum _{n=0}^{\infty } \int _{0}^{\infty }
q d q w_3\left (\frac{1}{\Delta } + \frac{1}{\Delta _1}\right )
\, .
\end{eqnarray}
Putting all expressions together one can write down (\ref{d18}) explicitely.
This general expression is a very bulky one, however, it can
be considerably simplified in two important limiting cases.

\subsection{VdW interactions}
\label{VdW}

We term VdW interaction the case of small thicknesses, i.e. $l \ll \lambda ^*$,
where $\lambda ^*$ characterizes the main absorption band of the cholesteric film.
Typically in known cholesteric liquid crystals \cite{CH92}, $\lambda ^* $ is of
the order of $10^{-4} cm$ and $q_0 \lambda ^* < 1$. Thus we can perform an expansion
over $q_0 l$ in (\ref{d7}) - (\ref{d18}) and get a relatively compact
answer for this limit
\begin{eqnarray}
\label{d19}
\mu (l)  = \mu _0(l) + a\frac{q_0^2}{l}
\, ,
\end{eqnarray}
where $\mu _0 (l) \propto l^{-3}$ is known \cite{DL61} for isotropic films VdW contribution, and
\begin{eqnarray}
\label{d20}
a  = \frac{1}{64\pi ^2 }\int _{0}^{\infty } dp \int _{0}^{\infty } d\xi 
\frac{\epsilon _a(i\xi )}{\epsilon _3(i \xi )}
\left [\frac{(\epsilon _1(i\xi ) + \epsilon _3 (i\xi ))(\epsilon _3 (i\xi ) + 1)}
{\epsilon _3(i \xi ) - \epsilon _2(i \xi ))(\epsilon _3(i \xi ) - 1)}\exp (x) - 1 \right ]^{-1}
\, .
\end{eqnarray}
Let us emphasize in passing that this anisotropic VdW contribution scales as $1/l$,
and it is more long ranged than known fluctuational forces in isotropic
homogeneous materials. It comes from the interplay of the characteristic
lengths (film thickness and cholesteric pitch), and the extra factor $q_0^2$ in (\ref{d19})
provides correct dimensionality.
Note also that the sign of this contribution (\ref{d20}) does not depend on the
cholesteric spiral chirality (left or right) as one can certainly expect
but it does depend on dielectric properties and could be positive
or negative. In the latter case the planar director orientation
(cholesteric axis along $x$-axis) is always stable, and in the former
case this orientation is energetically unfavorable and could become
unstable at a certain thickness $l_{cr}$.
The critical thickness is determined by a competition
of VdW contribution (\ref{d20}) and stabilizing
planar orientation short range anchoring energy $W$
\begin{eqnarray}
\label{d211}
l_{cr} \propto \frac{W}{a q_0^2}
\, .
\end{eqnarray}
This transition is analogous to the Freedericksz transition known
in liquid crystals \cite{GP95}, \cite{CH92}, however,
while the Freedericksz transition in liquid crystals is driven by the
quadratic coupling between an external magnetic (or electric)
field and the director, in our case the transition is induced by internal
electromagnetic field fluctuations (VdW forces).

Some comments about the validity of (\ref{d211}) seem in order here.
The critical thickness (\ref{d211}) is determined by a competition
of the VdW (long range) contribution into the bulk chemical potential,
and phenomenological (basically short range) anchoring energy $W$.
Clearly the approach misses short range contributions into the bulk properties.
If the only forces between particles had a range of the order of molecular dimensions 
(short range forces), the corresponding contributions would decrease exponentially
with increasing distance, therefore assuming that at the scale (\ref{d211}) the VdW contribution
is dominating, we believe that $l_{cr}$ is much larger than all molecular scales.
This assumption is not always justifiable, and in this case, (\ref{d211}) does not maintain numerical accuracy, but
nevertheless is useful and instructive for gaining some qualitative
insight on interplay short and long range forces.

\subsection{Casimir force}
\label{cas}
We now turn to the opposite limiting case 
$l > \lambda ^*$ when electromagnetic interaction retardation
effects become relevant, and it is referred as Casimir limit. 
Closely following to \cite{DL61}
we can replace in this case all dielectric permeabilities entering the general 
formula (\ref{d18}) for $\mu (l)$ by their static values,
and after some algebra end up with
\begin{eqnarray}
\label{d21}
\mu (l)  = \mu _0(l) + b\frac{1}{l^4} \cos (2 q_0 l)
\, ,
\end{eqnarray}
where $\mu _0 \propto l^{-4}$ is conventional isotropic
Casimir contribution, and the coefficient $b$ reads as
\begin{eqnarray}
\label{d22}
b = -\frac{1}{32 \pi ^2}\int _{0}^{\infty } dx \int _{1}^{\infty } dp
\epsilon _a(0) \sqrt \epsilon _3 (0) \frac{x^3}{p^2}
\frac{\phi \exp (x)}{(\phi \exp (x) - 1)^2}
\, ,
\end{eqnarray}
where we designate
\begin{eqnarray}
\label{d23}
\phi   = \frac{(s_{10} + p\epsilon _1(0))(s_{20} + p)}{(s_{10} - p\epsilon _1(0))
(s_{20} - p)}
\, ,
\end{eqnarray}
and
\begin{eqnarray}
\label{d24}
s_{10} = \sqrt {\frac{\epsilon _1(0)}{\epsilon _3(0)} - 1 + p^2}
\, ; \, s_{20} = \sqrt {\frac{1}{\epsilon _3(0)} - 1 + p^2}
\, .
\end{eqnarray}
The Casimir limit (\ref{d21}) is particularly instructive,
because it shows oscillating with a thickness dependence of the chemical potential.
Since the both contributions (monotonic and oscillating ones) decay
as the same power of $l$, it is especially interesting to hypothesize
the case (not forbidden in principle although formally beyond
our first order perturbation theory approximation)
when it becomes negative.
Suffice it to say that such an oscillating function $\mu (l)$
leads to a film formation which is stable only in a certain range of
its thickness.
This kind of successive thinning or wetting - drying
transitions are known for layered smectic phases (see e.g. \cite{GE90},
\cite{GP99}, \cite{PO01}). We have shown that Casimir forces in cholesteric liquid crystals
can be responsible for similar phenomena.
Lacking sufficient data on dielectric permeability frequency dispersion
we can at present discuss only the general qualitative features
of Casimir forces in cholesteric liquid crystals.

Actually our picture is not entirely correct since
we have neglected all short range forces, and this relatively
weak oscillating Casimir contribution can be swamped by stronger
short range forces. However by their nature, the short range forces
have no connection with the effects under consideration which are
due to cholesteric long wavelength orientational modulations.
Short range contributions to the chemical potential are the same in cholesterics and in 
analogous homogeneous systems (e.g., nematics). Thus, even
in the case of dominating short range forces, the Casimir contribution
(\ref{d21}) could be disentangled in differential measurements.
We anticipate that the discussed above effects
of Casimir film instabilities will be observable and that understanding
of underlying mechanisms will be essential
to predict and to use these phenomena.

\section{Conclusion} 
\label{con}
In summary, in this paper we have calculated 
electromagnetic fluctuational contributions
into cholesteric liquid crystal film  chemical potential,
and found the oscillating with the film thickness contribution (\ref{d21}). 
Quite remarkably this result illustrates that the collective
nature of the Casimir forces suggests it to have a non-trivial
dependence on film thickness, and even the sign of the Casimir force is dependent
of material parameters and structure.

Our theoretical approach can be extended in several directions.
First, our main result (\ref{d21}) with evident modifications can be applied
to a smectic liquid crystal film, where its dielectric permeability
acquires an oscillating contribution owing to one dimensional
density modulation. Indeed, for a smectic film we should replace
(\ref{d4}) by
\begin{eqnarray}
\label{c1}
\epsilon _{yy} = \epsilon _{zz} = \epsilon _3
\, ; \, 
\epsilon _{xx} = (\epsilon _3 + \epsilon _a) + \left (\frac{\partial (\epsilon _3 + 
\epsilon _a)}{\partial \rho }\right ) \cos (q_{sm} x)
\, ,
\end{eqnarray}
where we chose $x$-axis as a normal to smectic layers,
and $q_{sm}$ is smectic density modulation wave vector.
Of course the macroscopic approach, we used in this paper,
is valid only when the corresponding modulation periods
are large with respect to atomic or molecular scales. It
is always the case for cholesteric liquid crystals (a typical pitch is about $(500 - 700) nm $), 
but usually not true for standard thermotropic smectics, where
their density modulated 
with a period in the range of $(2 - 5) nm $
\cite{CH92}, \cite{GP95}. Thus as an application
of our results to smectics, we have in mind long period
lyotropic smectic liquid crystals and some other 
lamellar (layered) membrane structures \cite{PO92}
where period can be much larger. 

One more comment of caution. The relation between the fluctuational force
and chemical potential 
(in fact the mechanical equilibrium condition  of the type
of (\ref{d6}) we have used for cholesteric films),
is valid for liquids (or uniform in density
cholesteric liquid crystals having no solid-like
elasticity). In smectics, like in solids, an elastic
layer compressibility modifies the equilibrium, and there
is no simple relation between electromagnetic fluctuation
force and film chemical potential. For these cases our theoretical predictions
should be confronted with direct force or (in anisotropic
cases) angular torque measurements.

On equal footing one can consider an apparently
unrelated problem of electromagnetic fluctuations in solid
long periodic magnetic structures.
The issue of Casimir forces
in materials with nontrivial magnetic susceptibility
has been discussed recently
\cite{KK02}, \cite{BR02}, \cite{MB02}.
To treat the magnetic case we have to include
magnetic permeabilities $\mu $ of all three media into the equation for
the Green function (\ref{d1}) and to modify the boundary conditions
(\ref{d5}). The latter ones are read as the continuity of
\begin{eqnarray}
\label{c2}
D_{zz} \, , \, \frac{1}{\mu }\frac{d D_{zz}}{d x} \, , \,
D_{yy} \, , \, \frac{\epsilon }{w^2}\frac{d D_{yy}}{d x}
\, ,
\end{eqnarray}
and besides $w$ functions are now $w^2 = \epsilon \mu \omega _n^2 + q^2$.
Calculations become more involved (see e.g. \cite{DP95}) but in the frame work of
perturbation theory are still feasible analytically.
However results and even signs of fluctuational contributions
depend on many unknown functions describing dielectric and magnetic
permeability frequency dispersions.
A more specific study might become appropriate should suitable
experimental results become available.

\subsection*{Acknowledgments} 
One of us (E.K.) acknowledges support from INTAS (under No. 01-0105) Grant.
 
\appendix
 
\section{}                                                                                                      

We take the $y$-axis along the vector $\bf q$ and
in the coordinate frame attached to the system (\ref{d2}) 
in self-evident notations reads
\begin{eqnarray}
\label{a1}
\left [ (\epsilon _{zz} \omega _n ^2 + q^2 ) - 
\frac{\partial ^2}{\partial x^2}\right ]D_{zz} + \epsilon _{zy} \omega _n^2 D_{yz}
= - 4 \pi \delta (x - x^\prime )
\, ,
\end{eqnarray}
\begin{eqnarray}
\label{a2}
\left [\epsilon _{yy} \omega _n ^2  - 
\frac{\partial ^2}{\partial x^2}\right ]D_{yy} + \epsilon _{yz} \omega _n^2 D_{zy}
+ i\frac{\partial }{\partial x}q D_{xy} = - 4 \pi \delta (x - x^\prime )
\, ,
\end{eqnarray}
\begin{eqnarray}
\label{a3}
(\epsilon _{xx} \omega _n ^2 + q^2 )D_{xx}  +
i q\frac{\partial }{\partial x}D_{yx} 
= - 4 \pi \delta (x - x^\prime )
\, ,
\end{eqnarray}
\begin{eqnarray}
\label{a4}
\left [(\epsilon _{zz} \omega _n ^2 + q^2 ) - 
\frac{\partial ^2}{\partial x^2}\right ]D_{zy} + \epsilon _{zy} \omega _n^2 D_{yy}
= 0
\, ,
\end{eqnarray}
\begin{eqnarray}
\label{a5}
\left [(\epsilon _{zz} \omega _n ^2 + q^2 ) - 
\frac{\partial ^2}{\partial x^2}\right ]D_{zx} = 0
\, ,
\end{eqnarray}
\begin{eqnarray}
\label{a6}
\left [\epsilon _{yy} \omega _n ^2  - 
\frac{\partial ^2}{\partial x^2}\right ]D_{yx} + + \epsilon _{yz} \omega _n^2
D_{zx} + i q \frac{\partial }{\partial x} D_{xx} = 0
\, ,
\end{eqnarray}
\begin{eqnarray}
\label{a7}
\left [\epsilon _{yy} \omega _n ^2 - 
\frac{\partial ^2}{\partial x^2}\right ]D_{yz} + i q \frac{\partial }{\partial x}
D_{xz} + \epsilon _{zy} \omega
_n^2 D_{zz}
= 0
\, ,
\end{eqnarray}
\begin{eqnarray}
\label{a8}
(\epsilon _{xx} \omega _n ^2 + q^2) D_{xz} + i q \frac{\partial }{\partial x}
D_{yz} 
= 0
\, ,
\end{eqnarray}
\begin{eqnarray}
\label{a9}
(\epsilon _{xx} \omega _n ^2 + q^2) D_{xy} + i q \frac{\partial }{\partial x}
D_{yy} 
= 0
\, .
\end{eqnarray}
From these equations in the first order over the small parameter $\epsilon _a /\epsilon _3$
we get two decoupled equations for $D_{zz}$ and $D_{xx}$, which are in the
only non-trivial region $0 < x < l$ 
\begin{eqnarray}
\label{a10}
\left [w_3^2   - 
\frac{\partial ^2}{\partial x^2} + \epsilon _a \omega _n^2 \cos (2q_0 x)\right ]D_{zz} 
= - 4 \pi \delta (x - x^\prime )
\, ,
\end{eqnarray}
and
\begin{eqnarray}
\label{a11}
\left [w_3^2   - 
\frac{\partial ^2}{\partial x^2} - \frac{\epsilon _a w_3^2}{\epsilon _3}
- \frac{\epsilon _a q^2}{\epsilon _3 w_3^2} \right ]D_{yy} = 0
\, ,
\end{eqnarray}
and from the solution
to these equations using two relations (\ref{d6}) one can find two others
Green functions 
$D_{xy}$ and $D_{xx}$. 
General solutions to the equations (\ref{a10}), (\ref{a11})
read as
\begin{eqnarray}
&&
\nonumber
D_{zz} = 
C_1\exp (w_3 x)\left [1 - \frac{\epsilon _a \omega _n^2}{4 q_0 w_3}\sin (2q_0x)
+ \frac{\epsilon _a \omega _n^2}{4 w_3(w_3^2 + q_0^2)}(w_3\cos (2q_0x) +
q_0\sin (2q_0x)) \right ] \\
&&
\nonumber
+C_2\exp (-w_3 x)\left [1 + \frac{\epsilon _a \omega _n^2}{4 q_0 w_3}\sin (2q_0x)
- \frac{\epsilon _a \omega _n^2}{4 w_3(w_3^2 + q_0^2)}(-w_3\cos (2q_0x) +
q_0\sin (2q_0x)) \right ] \\
&&
\label{a12}
- \frac{2\pi }{w_3}\exp (-w_3 |x - x^\prime |)\left [1 + \frac{\epsilon _a \omega _n^2}{4 q_0 
(w_3^2 + q_0^2)}(q_0\cos (2q_0x) + w_3\sin (2q_0x) \right ] 
\end{eqnarray}
and
\begin{eqnarray}
&&
\nonumber
D_{yy} = 
A_1\exp (w_3 x)
\left [1 + \frac{\epsilon _a w_3}{4 q_0 \epsilon _3}\sin (2q_0x)
- \frac{\epsilon _a w_3}{4\epsilon _3(q_0^2 + w_3^2)}(w_3\cos(2q_0x)
+ q_0\sin(2q_0x)) + \frac{\epsilon _a q^2}{4\epsilon _3 w_3^2}\right ] \\
&&
\nonumber
+A_2\exp (-w_3 x)
\nonumber
\left [1 - \frac{\epsilon _a w_3}{4 \epsilon _3 q_0}\sin (2q_0x)
+ \frac{\epsilon _a w_3}{4 \epsilon _3(w_3^2 + q_0^2)}(-w_3\cos (2q_0x) +
q_0\sin (2q_0x)) + \frac{\epsilon _a q^2}{4 \epsilon _3 w_3^2} \right ] \\
&&
\nonumber
- \frac{2\pi w_3}{\epsilon _3\omega _n^2}\exp (-w_3 |x - x^\prime |)
- \frac{\pi w_3^2\epsilon _a}{\epsilon _3^2\omega _n^2}\exp (w_3 x)t_+ 
\\
&&
\label{a13}
+ \frac{\pi q^2\epsilon _a}{\epsilon _3^2\omega _n^2}\exp (w_3 x)s_+ 
+ \frac{\pi w_3^2\epsilon _a}{\epsilon _3^2\omega _n^2}\exp (-w_3 x)p_+ 
-\frac{\pi q^2\epsilon _a}{\epsilon _3^2\omega _n^2}\exp (-w_3 x)m_+ 
\, , 
\end{eqnarray}
where $C_i \, , \, A_i$ are constants which should be found from
the $x=0$ and $x=l$ boundary conditions, and in (\ref{a13}) we introduce
shorthand notations
\begin{eqnarray}
&&
\nonumber
t_+ = \exp(w_3 x^\prime - 2 w_3 x)\frac{-w_3\cos (2q_0x) + q_0\sin (2q_0x)}{2(q_0^2 + w_3^2)} \, ; \,
s_+ = - \frac{1}{2w_3}\exp (w_3 x^\prime - 2 w_3) \, ; \,
\\
&&
\label{a14}
m_+ = \frac{1}{2w_3}\exp (-w_3 x^\prime + 2w_3x) \, ; \,
p_+ = \exp (w_3 x^\prime )\frac{1}{2q_0}\sin (2q_0x)
\, .
\end{eqnarray}


\end{document}